# Phonon dispersions throughout the iron spin crossover in ferropericlase

Michel L. Marcondes[1], Fawei Zheng[2], and Renata M. Wentzcovitch[1,3]

[1] *Department of Earth and Environmental Sciences, Columbia University, Lamont-Doherty Earth Observatory, Palisades, NY 10964*
[2] *Institute of Applied Physics and Computational Mathematics, Beijing, 100088, China*
[3] *Department of Applied Physics and Applied Mathematics, Columbia University, New York, NY 10027*

Ferropericlase (Fp), $(Mg_{1-x}Fe_x)O$, is the second most abundant phase in the Earth's lower mantle. At relevant pressure-temperature conditions, iron in Fp undergoes a high spin (HS), S=2, to low spin (LS), S=0, state change. The nature of this phenomenon is quite well understood now, but there are still basic questions regarding the structural stability and the existence of soft phonon modes during this iron state change. General theories exist to explain the volume reduction, the significant thermo-elastic anomalies, and the broad nature of this HS-LS crossover. These theories make extensive use of the quasi-harmonic approximation (QHA). Therefore, dynamical and structural stability is essential to their validity. Here, we investigate the vibrational spectrum of Fp throughout this spin-crossover using *ab initio* DFT+Usc calculations. We address vibrational modes associated with isolated and (2nd) nearest neighbor iron ions undergoing the HS-LS state change. As expected, acoustic modes of this solid solution are resilient while optical modes are the most affected. We show that there are no soft phonon modes across this HS-LS crossover, and Fp is dynamically stable at all relevant pressures.

## I. INTRODUCTION

Ferropericlase (Fp) is the second most abundant mineral phase in the Earth's lower mantle. It might be responsible for up to 25 vol% of this region, which is responsible for ~ 55 vol% of the Earth. It is a solid solution of MgO and FeO (($Mg_{1-x}$,$Fe_x$)O with $X_{Fe} < 0.5$) in a rocksalt-type (B1) structure. Properties of Fe-bearing MgO are fundamental for understanding properties and processes taking place in Earth's deep interior. Today it is well known that at high pressures, iron in MgO undergoes a spin-state change from a high spin (HS) with total spin S = 2 to a low spin (LS) with S = 0. The discovery of this phenomenon[1,2] has raised many questions about its physical nature and geophysical implications, and it has been a topic of extensive research. Currently, general theories exist to explain the broad nature of this state change (a crossover) and the volume reduction throughout the crossover[3]. Also, the large thermo-elastic anomalies throughout the crossover[4–7] have been explained[8,9]. Some geophysical consequences, e.g., the insensitivity of compressional velocities, $V_P$, to lateral (isobaric) temperature variations[10], or the increase in adiabatic temperature gradient[11] in the lower mantle have also been derived. These theories make extensive use of the quasi-harmonic approximation (QHA). Therefore, dynamical and structural stability are essential to their validity.

Although static properties of this state change have been extensively studied, the dynamical stability of the iron environment still needs closer inspection. Brillouin scattering experiments show a softening in $C_{11}$ and $C_{12}$ elastic coefficients[4,5], impulsive stimulated scattering shows an additional softening in $C_{44}$[6], and inelastic x-ray scattering (NRIXS) produced anomalies in $C_{12}$ and $C_{44}$ but not in $C_{11}$[7]. Several theoretical and experimental investigations have subsequently addressed these anomalies[9,12–15]. Although there are inconsistencies in the experimental data, it is clear that this spin-state change softens some of Fp's elastic coefficients. This softening has been explained [9] by invoking the significant volume reduction of ~ 8%[3,8] of the Fe-octahedron with this spin-state change. The question remains whether localized vibrational mode instabilities occur during this state change and whether they could also impact the shear elastic coefficient $C_{44}$.

While theoretical investigations usually attempt to reproduce experimental or geophysically relevant iron concentrations, for our purpose, it is more important to investigate Fp with low iron concentrations. In this case, radically different iron configurations with isolated and interacting irons can be investigated, and the vibrational properties derived. Furthermore, experimental[16,17] and theoretical[17] studies show that FeO undergoes phase transitions culminating in a metallic phase at ~ 70 GPa and 1,900 K, or at 120 GPa and 300 K[18]. Therefore, MgO and FeO end-members are unlikely to produce an isomorphous solid solution. Indeed, it has been shown that Fp with high iron concentration dissociates into an iron-rich phase and a phase with small iron concentrations[19].

Here, we investigate the vibrational properties of $(Mg_{1-x},Fe_x)O$ for $X_{Fe}$ ~3% and ~6%, henceforth 3Fp and 6Fp, respectively, for isolated and nearest (sub-lattice) neighbor iron configurations.  We obtain phonon dispersions of distinct HS, LS, and mixed spin (MS) configurations across the entire pressure range of the lower mantle and compare them in detail with phonon dispersions in MgO.

In the next section, we describe the computational details of the calculations. In section III, we present the main results, followed by a summary of our findings in section IV.

## II.  COMPUTATIONAL METHOD

DFT calculations based on the local spin density approximation (LSDA) and spin-polarized generalized gradient approximations (σ-GGA) predict an incorrect metallic ground state for Fp[3]. Thus, we performed calculations using the LSDA plus Hubbard U method (LDA+$U_{sc}$)[20] with U calculated self-consistently[21] and structurally consistent[22]. The Hubbard U parameters used here are volume and spin-state dependent and are the same as those previously published[23]. This method describes the spin crossover equations of state in iron-bearing phases[24–27] accurately. Structural optimizations are performed with damped variable cell shape molecular dynamics[28,29] as implemented in the Quantum ESPRESSO code[30].

We constructed a 2x2x2 simple cubic supercell with 64 atoms with one and two iron ions corresponding to concentrations $X_{Fe} = 0.03125$ (3Fp) and $X_{Fe} = 0.0625$ (6Fp). To understand the effect of iron-iron interaction on Fp properties, we constructed two different supercells for 6Fp: one with iron ions as 2nd neighbors (2nn) and another with iron as 11th neighbors (11nn). Figure 1 illustrates the supercell configurations used in this work. We used a plane-wave energy cut-off of 80 Ry and a shifted 2x2x2 **k**-mesh to sample electronic states throughout this supercell's Brillouin Zone (BZ). The energy vs. volume curve was fitted to a 3rd order finite strain equation of state [31]. For each structure we calculated phonon dispersions at 0, 60, 80, and 135 GPa using density functional perturbation theory + U (DFPT+U)[32]. The calculated supercell phonon dispersions were subsequently unfolded into the first BZ of the primitive B1 structure using the phonon unfolding method[33–35]. We adopted a plane-wave based unfolding procedure[36], which projects the supercell modes into plane-wave-like modes with wave-numbers **q** in the first BZ of the B1 structure. This procedure is useful to identify changes in phonon dispersions caused by heavy translational symmetry breaking, as is the case here. Each of these calculations was performed for the HS and LS states. For 6Fp, we also calculated dispersions for an MS state with one iron in the HS and another in the LS state (HS-LS).

## III. RESULTS

Ferrous iron ($Fe^{2+}$) has $3d^6$ electronic configuration and occupies an octahedral site with $O_h$ point group symmetry in Fp. The $O_h$ symmetry causes a crystal field splitting of the $d$-orbitals' energy-producing a doublet with $e_g$ and a triplet with $t_{2g}$ symmetry. $e_g$ and $t_{2g}$ orbitals have lobes pointing toward and away from the nearest oxygen neighbors, respectively. Accordingly, the $t_{2g}$ states have lower energy than the $e_g$ states. For iron with a $3d^6$ configuration, several orbital occupancies with different spin states are possible, each one with a slightly different energy level structure due to further Jahn-Teller-type symmetry breaking. For smaller crystal field splitting at low pressures, the spin-up orbitals have lower energy forming the HS state according to Hund's rule, with five electrons with spin up and one with spin down (see FIG. 2a). With increasing pressure, the crystal field splitting increases, and at soring pressures, only the $t_{2g}$ orbitals are occupied with two electrons each, forming the LS state with S=0 (see FIG. 2b). The intermediate spin (IS) state with four electrons with spin up and two with spin down and S=1 is not energetically competitive in any configuration in Fp[23].

Figure 3 shows electron density iso-surfaces for the majority and minority electrons in both spin states. In the HS state, the Jahn-Teller distortion breaks the octahedral symmetry and increases bond-lengths by ~2% in the (x,y) plane and by 1% in the z-direction w.r.t. of that of MgO. In the LS state, there is no Jahn-Teller distortion, and all bond-lengths decrease by ~1%. Thus, there is an overall decrease of ~8% in the octahedral volume across the spin-state change. This effect has been shown to produce elastic anomalies in the bulk modulus [8], $K_T$ and $K_S$ alike, and in $C_{11}$ and $C_{12}$ [5]. Whether vibrational instabilities throughout the spin-crossover could produce elastic softening also in $C_{44}$ has been a subject of debate.

HS and LS Fp, even for the low iron concentrations investigated here, have different compressibilities as reflected in the equation of state (EOS) parameters given in TABLE 1. For 6Fp, TABLE 1 also shows EOS parameters for the MS state. For 3Fp, the volume change $\Delta V_0$ is ~ 0.7% across the spin-state change. For 6Fp with iron in the 11nn configurations (FIG 1c) $\Delta V_0$ is

~1.4%, i.e., twice the difference in 3Fp. For 6Fp with iron ions starting in the MS state and only one iron undergoing a spin-state change, $\Delta V_0$ is also ~0.7%. These results support the quasi-ideal solid solution model used to investigate the spin crossover in Fp[3,8–10,23]. In contrast, for 6Fp with iron ions in the 2nn configuration, $\Delta V_0$ is ~1.1% with both irons undergoing the spin change, and $\Delta V_0$ is ~ 0.6% from the MS state, indicating a non-negligible iron-iron interaction. Nevertheless, the relative abundance of these 11nn and 2nn configurations is related to the enthalpy differences. As shown below (see FIG. 4), the 11nn configuration is more stable and should be more abundant than the 2nn configuration, alleviating in part the non-ideality problem.

For a fixed $X_{Fe}$ and atomic configuration, the bulk moduli for different spin states vary according to expectations based on the behavior of the 2nd order finite strain equation of state where $P/K_0$ is a unique function of $V/V_0$[37,38]: the larger the volume, the smaller the bulk modulus. This is not the case for the same spin state but different iron configurations or concentrations, indicating non-negligible iron-iron interaction, mostly of elastic nature. However, configurations with closer iron ions have higher enthalpy (see FIG. 4) and should be less abundant.

In the present calculation, the spin-state change occurs when the static enthalpy difference ($\Delta H$) between the HS and LS state vanishes:

$$\Delta H = \Delta E + P\Delta V = 0$$

where $\Delta E$ and $\Delta V$ are the differences between total internal static energies and volumes, respectively, and $P$ is the static pressure. Figure 4 shows $\Delta H$ for all spin-state changes, configurations, and concentrations studied here. The red baseline corresponds to the enthalpy of the HS state, and, as a reference, it is the same for 3Fp and 6Fp. The green lines represent enthalpy differences between MS and HS states of 6Fp with 11nn and 2nn configurations. As indicated in the inset, the first transition is from the HS-HS to the HS-LS state, irrespective of atomic configuration. The HS to MS transition in the 11nn configuration having larger $\Delta V$ transitions at a slightly lower pressure $P_T$ (~-1 GPa) than the 2nn configuration. This transition in 6Fp is followed by a second spin-state change from the HS-LS to the LS-LS state (crossing of green and pink/magenta lines). These transition pressures differ by ~5 GPa, irrespective of ionic configuration, as indicated in the inset. Therefore, atomic and spin configurations change the $P_T$ due to iron-iron interaction. In 3Fp, the spin-state change occurs at intermediate pressures (crossing of blue and red lines). All these state changes occur within a range of ~6 GPa. Previous calculations have indicated that within obtainable accuracy $P_T$ is independent of $X_{Fe}$ up to $X_{Fe} \sim 18\%$ [3,39]. Our results for low iron concentrations confirm that iron-iron interaction does not change $P_T$ significantly but can increase the spin crossover pressure range, $\Delta P_T$, as seen experimentally [40,41] and theoretically verified in $(Mg_{1-x},Fe_x)SiO_3$[42] and $(Mg_{1-x},Fe_x)(Si_{1-x},Fe_x)O_3$[43]. For $X_{Fe} > 18\%$ not only $\Delta P_T$ increases but the mid-point crossover pressure can also increase due to iron-iron interaction [44,45]. Here we obtained a transition pressure of 72 GPa, in good agreement with a previous calculation [23,45] but higher than experimental values that can vary from 40 to 70 GPa [40,44,46] with iron concentrations. The inclusion of vibrational effects should increase slightly the $P_T$ values [8]. For an early review of the theoretical and experimental values of $P_T$, see Ref. [47].

Phonon calculations assume a quadratic behavior of the energy to small ionic displacements. To confirm that this is the case during the spin-state change, we computed energy vs. small and isotropic changes in the octahedral Fe-O bond-lengths, as shown in FIG. 5 for several pressures. For both spin states, $E$ vs. $\Delta l$ curves fit well a parabolic function confirming this quadratic dependence throughout the entire pressure range of the Earth's lower mantle (up to 135 GPa). Besides, FIG. 5 indicates that both states are highly stable at all relevant pressures with the HS (red reference curve) state being more stable at 0 GPa and the LS state more stable at high pressures (80 and 135 GPa).

It has already been shown that calculated acoustic velocities in Fp[8,9] reproduce the elastic anomalies observed during the spin-state change [4,5]. The origin of these anomalies was attributed to the abnormal volume reduction in the compression curve caused by the spin-state change, not to vibrational instabilities. Figure 6 shows phonon dispersions along the high symmetry lines of the B1-type phase BZ for 3Fp in the HS and LS states. These dispersions were obtained unfolding[35,36] supercell phonon dispersions into the larger BZ of the B1-type structure (FIG. 1). For comparison, we also plot phonon dispersions of pure MgO periclase. These supercell calculations impose an artificial iron ordering but offer a realistic glimpse of the effect of iron alloying on the vibrational frequencies of MgO. Phonon modes of the 64-atom supercell were projected into multiple plane-wave modes with wave-number $\mathbf{q}$'s in the first BZ of B1-type structure. These projection amplitudes squared add to a value smaller than one, therefore the ghostly nature of these dispersions which track closely the phonon dispersions of pristine MgO (black dashed lines). The darker (lighter) points represent modes with polarizations resembling closely (differing strongly from) the primitive cell polarization modes with small (large) changes caused by translational symmetry breaking. The acoustic mode dispersions close to the $\Gamma$ point are resilient under translational symmetry breaking with acoustic velocity changes caused primarily by alloying density changes. At the $\Gamma$ point, transverse and longitudinal optical (TO and LO) mode frequencies are very similar to those of MgO with no significant new splitting introduced by translational symmetry breaking. This result indicates that vibrational modes associated with the iron impurity are resonant with those of the host crystal[48,49]. At 0 GPa, the TO and LO frequencies have energies of 50.8 meV and 88.0 meV, respectively, which are in good agreement with experimental values of 48.2 meV and 88.9 meV[13]. Throughout the entire pressure range of the lower mantle, both spin states have stable phonons. No phonon softening is observed with the main difference with respect to the dispersion in pristine MgO being the optical branch dispersions. This effect is expected on the basis of the artificially periodic supercell symmetry reduction. In summary, the low-pressure HS state remains metastable above ~70 GPa up to 135 GPa, and the high-pressure LS state remains metastable below ~70 GPa down to 0 GPa. These results validate the use of the QHA for free energy computations throughout the spin state change.

Although there is no phonon softening for isolated iron ions, the question remains whether phonons are still stable when iron-iron interaction is not negligible. Figure 7 shows phonon dispersions for 6Fp in the 11nn and 2nn configurations near $P_T$, i.e., at 60 and 80 GPa. Since the symmetry is more heavily broken in this case than in 3Fp, unfolded phonon dispersions are less well defined. Three spin state cases are shown, i.e., HS, LS, and MS (LS-HS). All these states have similar mode frequencies with seemingly small and hardly definable changes caused by ionic arrangements in the supercell. Again, the acoustic dispersion close to the $\Gamma$ point is not too different from those of pristine MgO. For all cases, the significant changes are in the optical mode dispersions, similarly to the case of 3Fp.

For solid solutions, the vibrational spectrum is more concretely depicted by the partial vibrational density of states (PVDOS). Figure 8 shows the PVDOS of 6Fp in the HS and LS states with iron ions in the 2nn and 11nn configurations. As shown in FIG. 5b, the curvature of the energy vs. bond-length line of the HS state for an isolated iron in 3Fp is larger than that for the LS state, a result that also applies to 6Fp in the 11nn configuration. The optical modes with the highest frequencies involve oxygen displacements primarily. In the 6Fp 11nn configuration, the Fe-O bonds are longer than in MgO and cause a compressive stress field in the surrounding oxygen (and iron) ions that increase the highest optical mode frequencies slightly (see FIG. 8c). The opposite occurs in the LS 11nn configuration with smaller Fe-O bond-lengths, which "decompress" the nearest neighbor oxygen environment and reduce the highest optical mode frequency (see FIG. 8b). A similar relationship between the highest frequencies in HS 2nn and LS 2nn configurations also holds. The maximum frequency in the HS 11nn and 2nn configurations indicate that in the 2nn configuration, the stress field around iron ions is smaller in the 2nn configuration, which is supported by the smaller equilibrium volume of this configuration. A similar relationship also holds between the highest mode frequencies of the LS 11nn and 2nn configurations. Although not too surprising, these results seem to cast doubt on the assumptions made by the vibrational virtual crystal model (VVCM) developed to investigate the thermodynamic properties of Fp[50], which motivates further research on the thermodynamic properties of Fp using more realistic VDOS and multi-configurations. Nevertheless, The volume reduction effect across the spin state change is a first order effect properly accounted for by the VVCM.

The main effect of iron alloying is a reduction in the contribution of magnesium displacement modes at low frequencies caused by an increased weight of iron displacement in these modes. In HS and LS states alike, modes involving iron displacements are slightly more localized in the 2nn configuration compared to those in the 11n configuration.

Given that there are no imaginary frequencies even in the worst case of the strongest iron-iron interaction, i.e., 2nn configuration in 6Fp, it seems safe to assume that phonon instabilities should not occur either for relevant iron concentrations ($X_{Fe} < 20\%$) throughout the entire pressure range of Earth's mantle.

## IV. SUMMARY

We have investigated the vibrational spectrum of ferropericlase (Fp), $(Mg_{1-x}Fe_x)O$, for $X_{Fe} \sim 3\%$ and $\sim 6\%$. Although the transition pressure is known to increase with increasing $X_{Fe}$ for $X_{Fe} > 25\%$[44,47], the spin transition pressure $P_T$ for our solid solutions with relatively small $X_{Fe}$ is independent of $X_{Fe}$. We see, however, a small dependence on $X_{Fe}$ of the crossover pressure range $\Delta P_T$ due to iron-iron interaction, with different iron configurations undergoing a spin-state change at slightly different pressures. We showed that iron-iron (mostly elastic) interaction increases the transition pressure, as observed experimentally for larger $X_{Fe}$. Both HS and LS states are dynamically/vibrationally stable throughout the whole pressure range of the Earth's lower mantle (up to 135 GPa), with no soft phonons or imaginary frequencies appearing near the spin-state change. These results fully validate the use of the quasiharmonic approximation (QHA) in previous free energy calculations of the Fp spin-crossover phase diagram[3,8] and elastic properties[8,9].

As indicated in previous works[8,9], the observed elastic softening is caused by the anomalous volume reduction produced by the spin-state change and the continuous change in the spin population in the mixed spin state. This phenomenon affects the bulk modulus[8] as well as the diagonal and off-diagonal elastic coefficients, but not the shear ones[9]. As previously indicated[15], the observed elastic softening in $C_{44}$ in some experiments appears to be caused by extrinsic effects.

These results confirm that the "localized" pressure-induced spin-state change in iron in the octahedral site resembles a first-order phase transition. Both states stable across the transition pressure separated by an energy barrier that stabilizes both phases and often causes hysteresis in high-pressure experiments. For low ($X_{Fe} < 20\%$) or high ($X_{Fe} > 20\%$) iron concentrations, iron-iron interaction can broaden the transition pressure range even at 0 K. This effect is distinct from and superposes to the well-known spin-crossover broadening effect caused electronic/magnetic configuration entropy at high temperatures[3,8].

*Acknowledgments* - This work was funded in part by National Science Foundation awards EAR-1503084 and EAR-1918126 (M. L. M.) and in part by the US Department of Energy award DESC0019759 (R.M.W.). This work used the Extreme Science and Engineering Discovery Environment (XSEDE), USA, which was supported by the National Science Foundation, USA Grant Number ACI-1548562. Computations were performed on Stampede2, the flagship supercomputer at the Texas Advanced Computing Center (TACC), The University of Texas at Austin generously funded by the National Science Foundation (NSF) through award ACI-1134872.

TABLE 1. Equation of state parameters obtained in static LSDA+Usc calculations in Fp.

| $X_{Fe}$ | HS | | | LS | | | MS (1 HS & 1LS) | | |
|---|---|---|---|---|---|---|---|---|---|
| | $V_0$ (Å$^3$) | $K_0$ (GPa) | $K_0'$ | $V_0$ (Å$^3$) | $K_0$ (GPa) | $K_0'$ | $V_0$ (Å$^3$) | $K_0$ (GPa) | $K_0'$ |
| 0.03125 | 74.19 | 170.3 | 4.15 | 73.66 | 172.1 | 4.15 | – | – | – |
| 0.0625 (11nn) | 74.38 | 171.0 | 4.16 | 73.32 | 174.8 | 4.17 | 73.86 | 171.3 | 4.23 |
| 0.0625 (2nn) | 74.15 | 172.6 | 4.05 | 73.34 | 174.6 | 4.17 | 73.85 | 173.0 | 4.16 |

**FIGURE CAPTIONS**

FIG. 1. Supercell structures used in the calculations of phonon dispersions. Red, orange, and blue spheres represent oxygen, magnesium, and iron, respectively.

FIG. 2. Predominant $d$ level splitting diagrams for $Fe^{2+}$ in an octahedral site a) in the high spin (HS) state and b) in the low spin (LS) state.

FIG. 3. Electron density iso-surfaces for the a) majority and b) minority electrons in the HS and in the c) LS states in iron in the $3d^6$ electronic configuration.

FIG. 4. Enthalpy difference between HS and LS states for $Mg_xFe_{1-x}O$. The transition pressure is ~70 ($\pm\mathbf{3}$) GPa in this calculation.

FIG. 5. Total energy vs Fe-O bond-length change in Fp with $\boldsymbol{X_{Fe} = 3}$% showing a highly quadratic behavior. The HS state at 0 GPa is used as a reference.

FIG. 6. Phonon dispersions in 3Fp for the atomic configuration shown in FIG. 1a (64 atoms supercell) unfolded in the first Brillouin zone of the face-centered primitive cell of MgO. These dispersions are compared with the phonon dispersion of MgO (dashed black lines). There are no phonon instabilities in the low-pressure HS state up to 135 GPa or in the high-pressure LS state down to 0 GPa.

FIG. 7. 6Fp phonon dispersion for HS, LS, and MS (HS-LS) states in the 2nn (FIG. 1b) and 11nn (FIG. 1c) configurations unfolded into the BZ of the B1-type structure.

FIG. 8. Total and partial vibrational density of states of 6Fp for all configurations investigated compared to that of pristine MgO at 60 GPa.

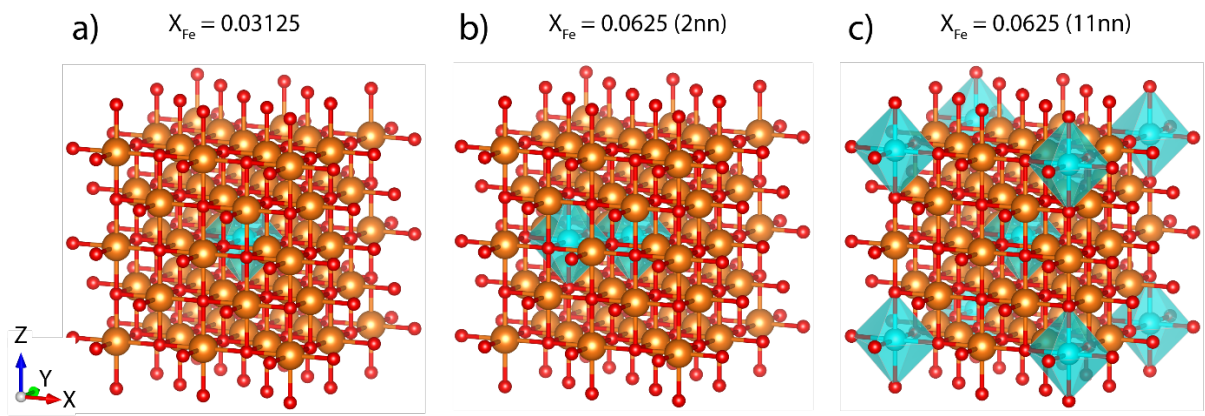

a) $X_{Fe} = 0.03125$  b) $X_{Fe} = 0.0625$ (2nn)  c) $X_{Fe} = 0.0625$ (11nn)

**Figure 1**

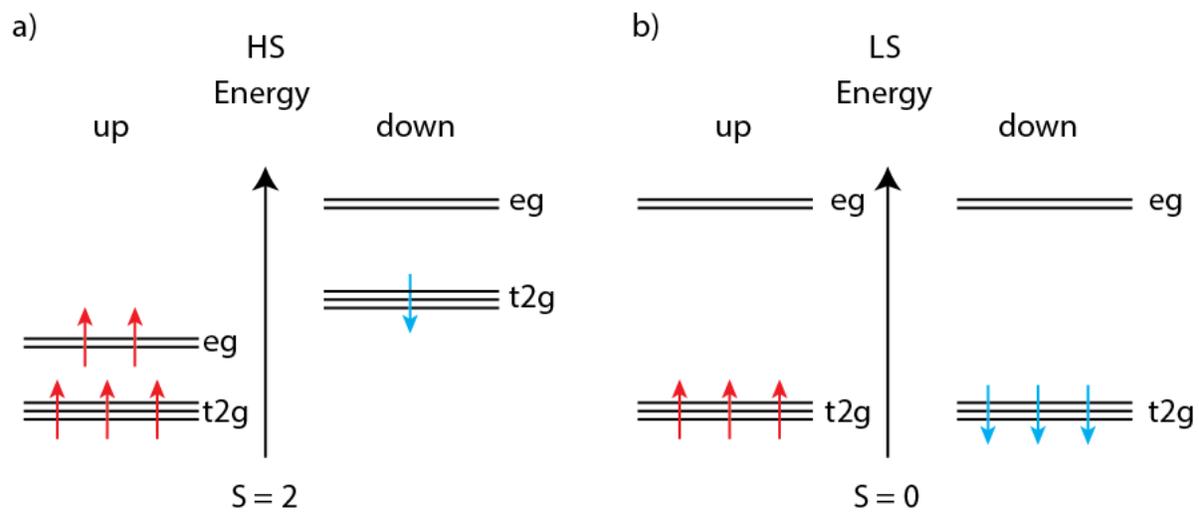

**Figure 2**

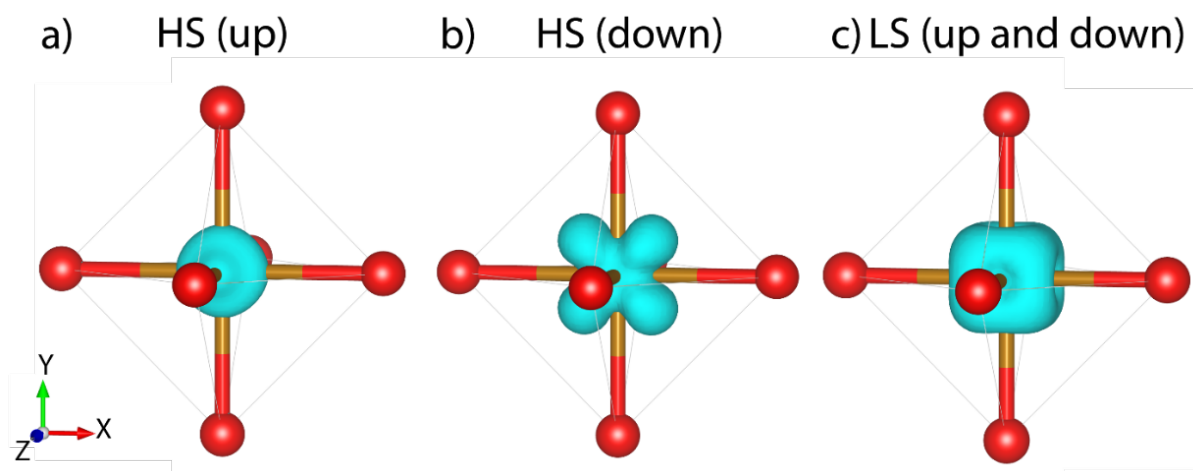

**Figure 3**

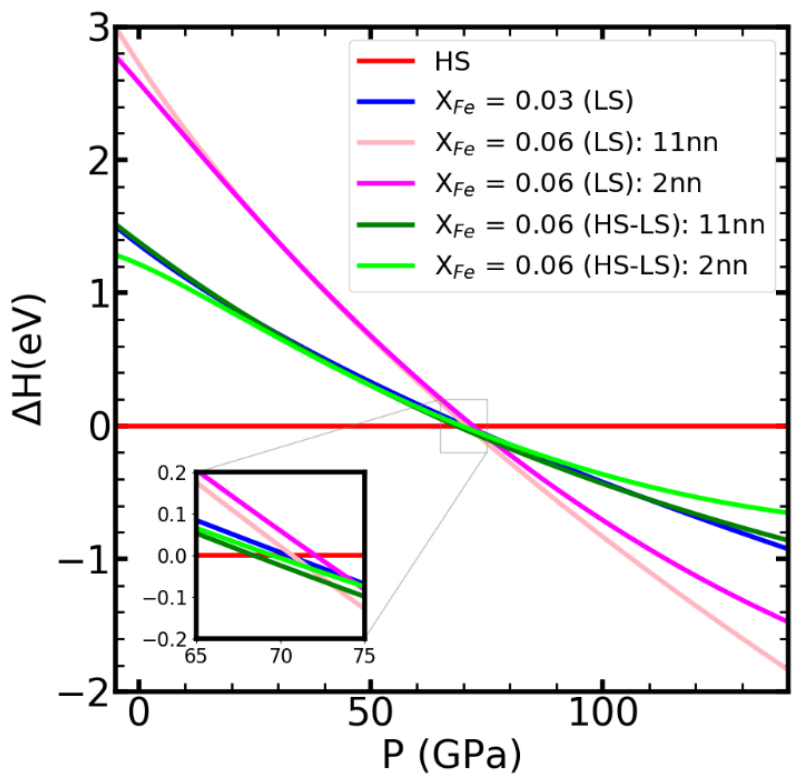

**Figure 4**

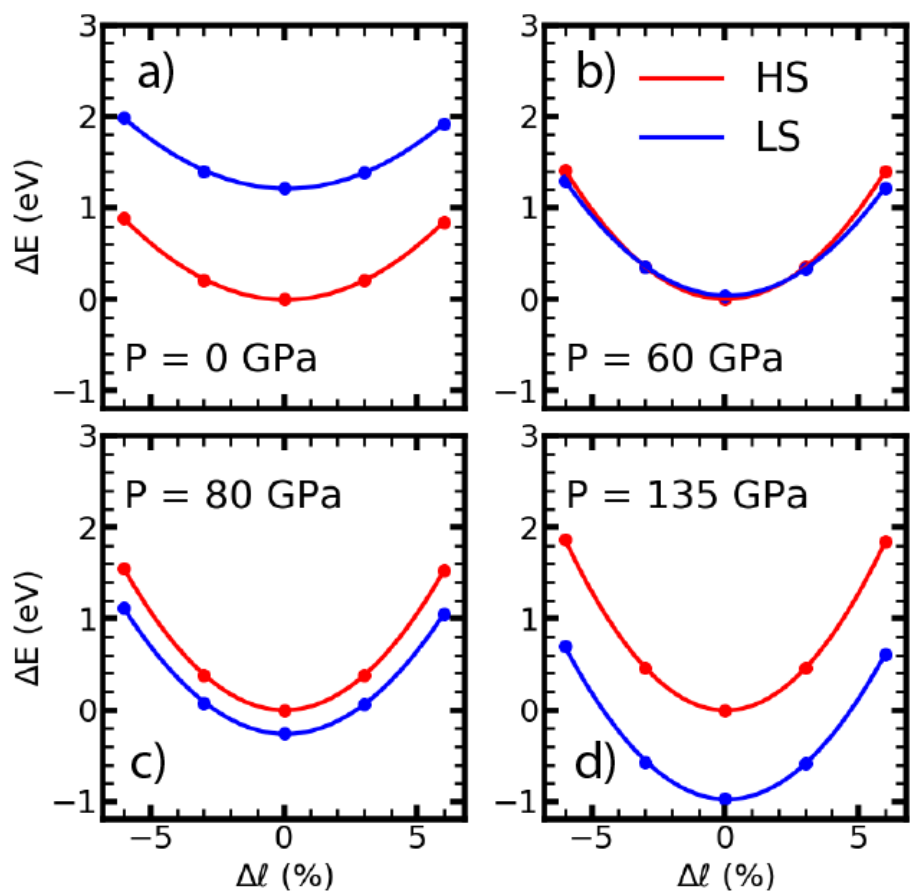

**Figure 5**

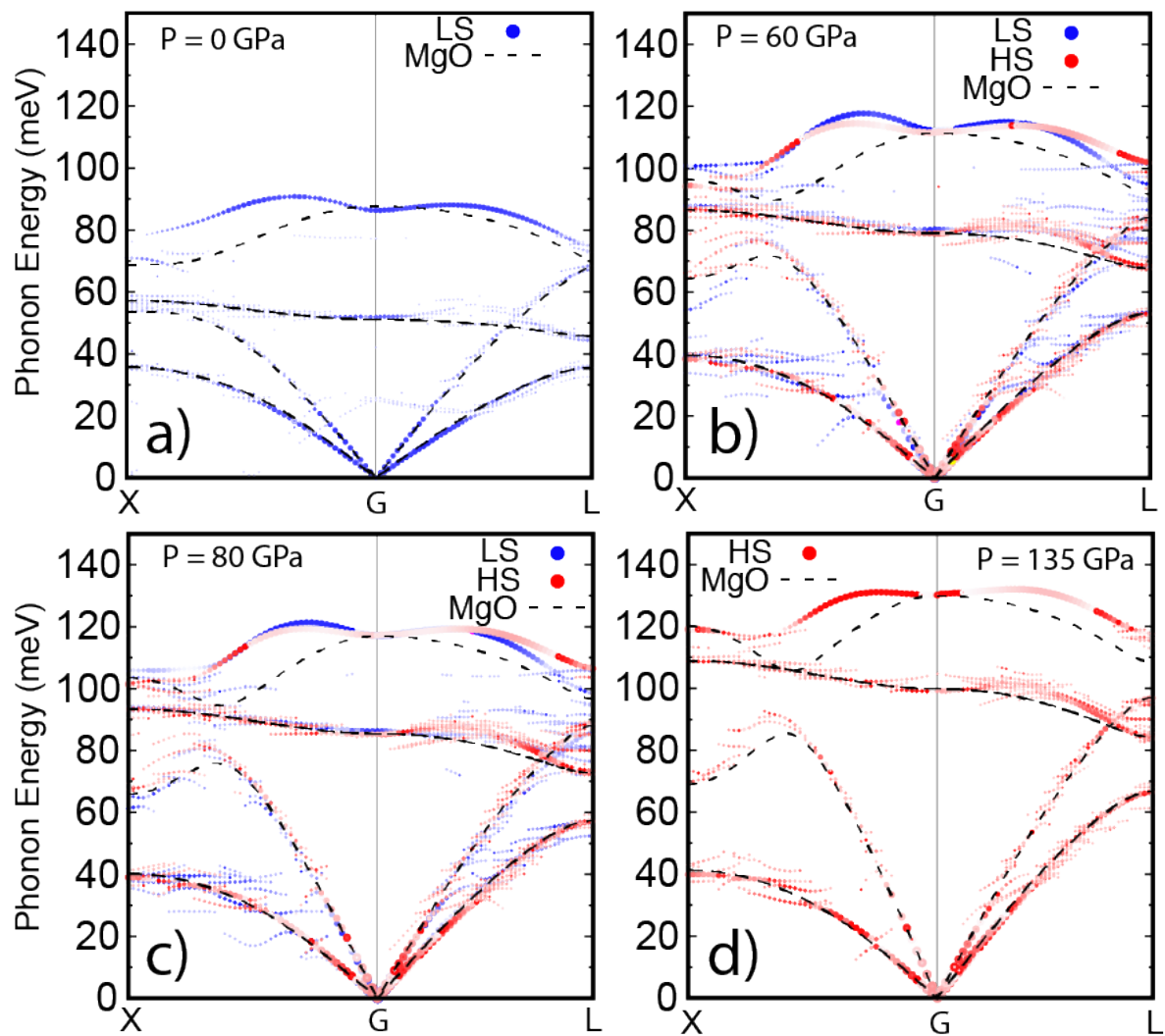

**Figure 6**

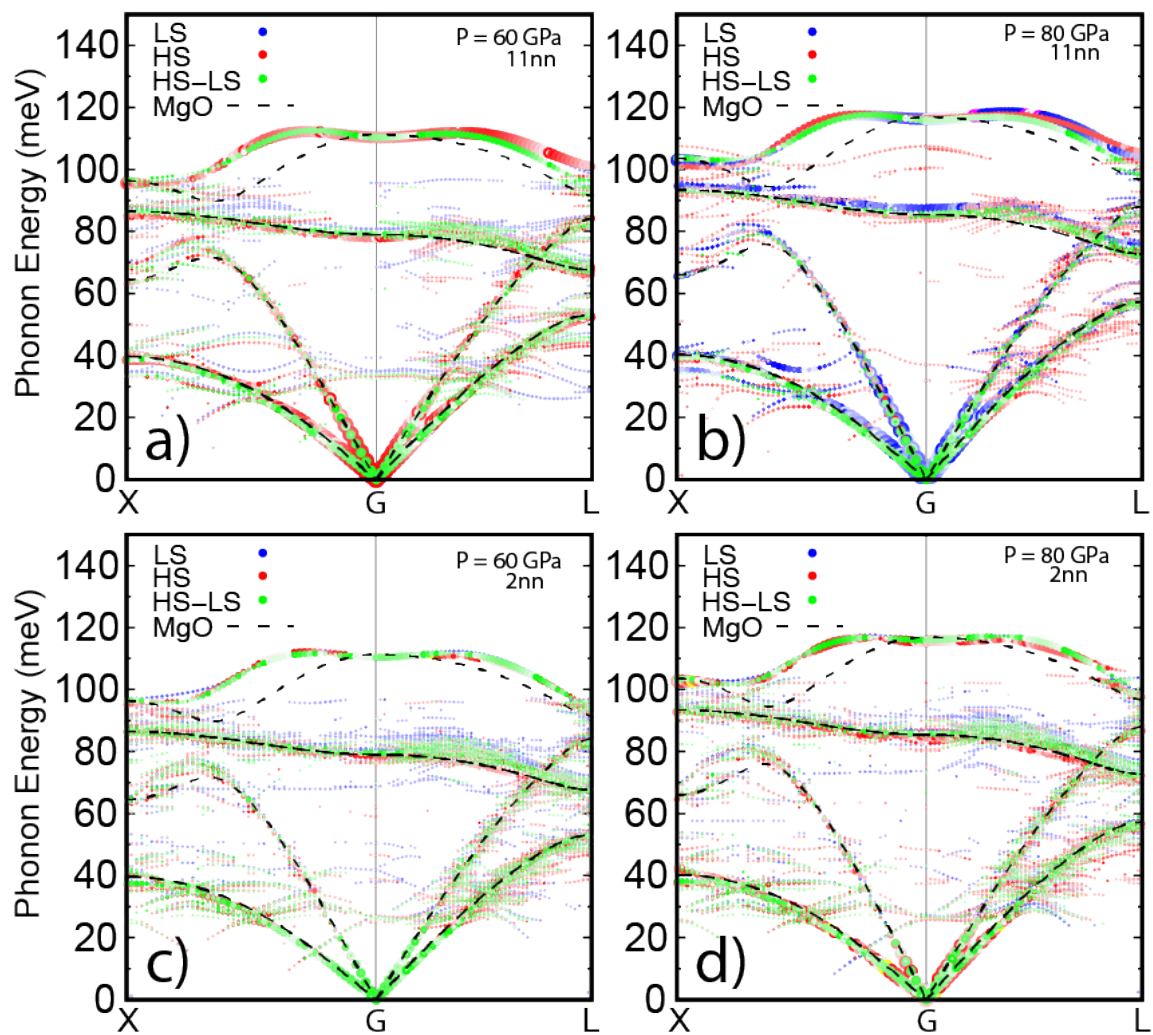

**Figure 7**

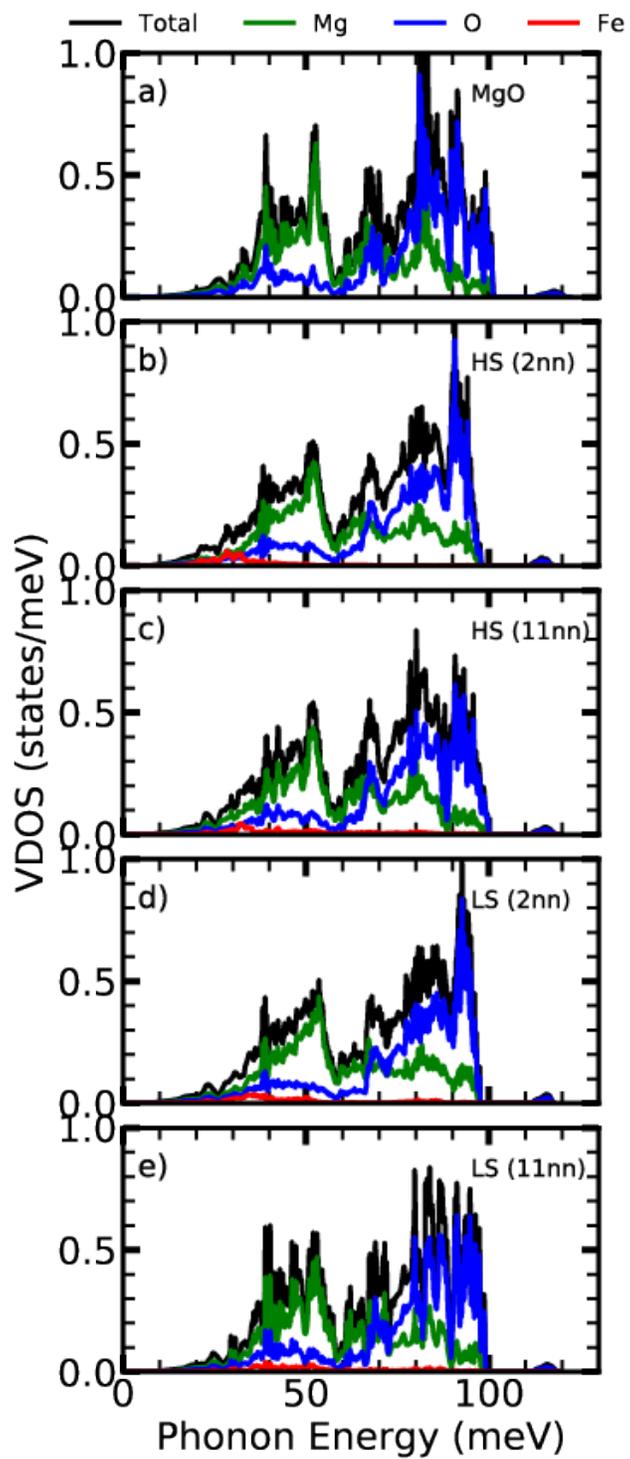

**Figure 8**